\documentclass[aps,prb,twocolumn,showpacs,groupedaddress]{revtex4}

\usepackage{amssymb}   
\usepackage{graphicx}
\usepackage{dcolumn}
\usepackage{bm}
\usepackage{amssymb} 

\begin{document}

\title{Crystalline electric fields and the ground state
of Ce$_{3}$Rh$_{4}$Pb$_{13}$.}

\author{D. A. Sokolov$^{1}$, M. C. Aronson$^{1}$, C. Henderson$^{2}$, and J. W. Kampf$^{3}$}

\address{$^{1)}$ Department of Physics, University of Michigan, 450 Church Street, Ann Arbor, MI 48109-1040,
USA\\}
\address{$^{2)}$ Department of Geological Sciences, University of Michigan, 1100 North University, Ann Arbor, MI 48109-1005,
USA\\}
\address{$^{3)}$ Department of Chemistry, University of Michigan, 930 North University, Ann Arbor, MI 48109-1055,
USA\\}

\begin{abstract}
{We have succeeded in synthesizing of single crystals of a new
intermetallic compound, Ce$_{3}$Rh$_{4}$Pb$_{13}$. Magnetic
susceptibility measurements indicate that the Ce moments are highly
localized, despite the metallic character of the electrical
resistivity.  Heat capacity measurements reveal that the cubic
crystal electric field lifts the six-fold degeneracy of the
Ce$^{3+}$ ground state, with the quartet state separated by
approximately 60 K from the doublet ground state. The magnetic field
dependence of the heat capacity at low temperature indicates a
further splitting of the doublet, but no sign of magnetic order was
found above 0.35 K.}

\end{abstract}
 \maketitle

Rare-earth based intermetallic compounds continue to attract a great
deal of interest due to the richness of the magnetic and electronic
phenomena observed in these systems~\cite{stewart2001,onuki2004}. In
particular, non-Fermi liquid behavior was found in the vicinity of a
zero temperature magnetic phase transitions, often manifested by
power-law temperature dependences of the magnetic susceptibility and
logarithmic temperature dependence of the heat capacity divided by
the temperature. Synthesis of novel rare-earth intermetallics in a
pure form, typically achieved by growing single crystals, is
instrumental in identifying new physics associated with the magnetic
ordering at zero temperature.

The compounds with a general formula R$_{3}$Rh$_{4}$Sn$_{13}$, where
R is a rare-earth element, were first synthesized by Remeika
\emph{et al}.~\cite{remeika1980}. It was found that the materials
with R=Tm, Lu, Y were superconductors, while compounds with R=Tb,
Dy, Ho ordered magnetically. The R=Eu member of this family
demonstrated reentrant superconductivity, in turn destroyed by the
onset of ferromagnetic order. Subsequently, other members of a
broader family R$_{3}$T$_{4}$X$_{13}$, where T=transitional metal
and X is metalloid from 4th or 5th group, were studied and found to
order antiferromagnetically at low
temperatures~\cite{takayanagi1994,aoki1996,nagoshi2006}.
Interestingly, heavy fermion Ce$_{3}$Ir$_{4}$Sn$_{13}$ orders
antiferromagnetically and demonstrates an intriguing double peak
feature near ordering temperature on the temperature dependence of
the heat capacity~\cite{takayanagi1994,nagoshi2005}. In heavy
fermion Ce$_{3}$Pt$_{4}$In$_{13}$, the Kondo and RKKY energy scales
are nearly equivalent and their balance can be changed by the
applied pressure~\cite{hundley2001}. Recently, new
Ln$_{3}$Co$_{4}$Sn$_{13}$ compounds, with Ln=Ce,La were
reported~\cite{thomas2006, cornelius2006}. Ce$_{3}$Co$_{4}$Sn$_{13}$
is a heavy fermion, which does not order magnetically above
T$\sim$0.35 K, while its La counterpart is a conventional
superconductor.

Such a panorama of magnetic properties and the variety of electronic
ground states in cubic R$_{3}$T$_{4}$X$_{13}$ motivated us to
synthesize a new compound with T=Rh, and X=Pb. We report here a
systematic study of the physical properties of a novel cubic
metallic compound Ce$_{3}$Rh$_{4}$Pb$_{13}$, which we carried out in
a search for new materials lying close to the quantum critical
point.

Single crystals of Ce$_{3}$Rh$_{4}$Pb$_{13}$ were grown from Pb
flux. Single crystal x-ray diffraction experiments were carried out
at room temperature using a Bruker 1K CCD based single crystal
diffractometer with Mo K$\alpha$ radiation. The structure was solved
and refined using SHELXTL package~\cite{sheldrick2001}. Electron
microprobe measurements were performed using a Cameca SX100
microprobe spectrometer calibrated with elemental Ce, Rh, and Pb
standards. The magnetic susceptibility was measured using a Quantum
Design Magnetic Phenomenon Measurement System at temperatures from
1.8 K to 300 K. The electrical resistivity $\rho$ of
Ce$_{3}$Rh$_{4}$Pb$_{13}$ was measured by the conventional
four-probe method between 0.35 K and 300 K in zero magnetic field.
Measurements of the heat capacity were performed using a Quantum
Design Physical Property Measurement System at temperatures from
0.35 K to 70 K and in magnetic fields up to 2 T.

Electron microprobe experiments found that the stoichiometry was
uniform over the surface of the crystals, with the elemental ratios
for Ce:Rh:Pb of 3$\pm0.02:4\pm0.04:13\pm$0.02. Single crystal X-ray
diffraction experiments confirmed this stoichiometry and found a
cubic structure (space group Pm$\overline{3}$n, No 223) with
a=10.0010(6) {$\AA$}, which is shown in Fig.~1. The unit cell
contains 2 formula units. Corrections were made for absorption and
extinction, and the atomic positions were refined with anisotropic
displacement parameters. The results of the X-ray diffraction
experiments are summarized in Tables~I,II.

\begin{table}
\caption{\label{tab:table2}Structural parameters of
Ce$_{3}$Rh$_{4}$Pb$_{13}$ at 293 K. Space group Pm-3n;
a=10.0010(6){$\AA$}, V=1000.30(10) {$\AA$}$^{3}$, calculated density
11.705 g/cm. Agreement indices: R$_{1}$=0.0398, wR$_{2}$=0.1158,
goodness of fit 1.256.}
\begin{ruledtabular}
\begin{tabular}{cccccccc}
Atom& Site &x & y &
 & z & U$_{eq}$\footnotemark[1] ({$\AA$}$^{2}$)\\
\hline Pb(1)& 2a & 0 & 0 & \
&0 & 0.027(1) \\
Pb(2)& 24k & 0.1557(1) & 0.3070(1) &
& 0 & 0.024(1)\\
Ce& 6d & 0 & 0.5 &
& 0.25 & 0.023(1)\\
Rh& 8e & 0.25 & 0.25 &
& 0.25 & 0.020(1) &\\
\end{tabular}
\end{ruledtabular}
\footnotetext[1]{U$_{eq}$ is defined as one-third of the trace of
the orthogonalized U$_{ij}$ tensor.}
\end{table}

\begin{table}
\caption{\label{tab:table2}Selected bond lengths ({$\AA$}) in
Ce$_{3}$Rh$_{4}$Pb$_{13}$ at 293 K.}
\begin{ruledtabular}
\begin{tabular}{cccccccc}
\hline Ce-Pb(1)& 5.591(1) & Ce-Pb(2)($\times$8)& 3.5220(7)\\
Ce-Pb(2)($\times$4)& 3.4897(9) & Ce-Rh& 3.5359(2) & \\
Rh-Pb(2)& 2.7321(4) & Rh-Pb(1)& 4.331(1)& \\
Pb(1)-Pb(2)& 3.4423(10) & Pb(2)-Pb(2)& 3.3182(15)& \\
Pb(1)-Pb(1)& 8.661(1)&\\
\end{tabular}
\end{ruledtabular}
\end{table}

\begin{figure}
\includegraphics[scale=0.5]{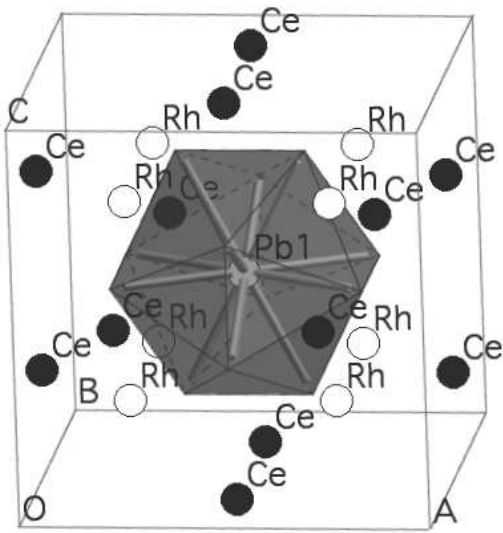}
\caption{\label{fig:epsart} Crystal structure of
Ce$_{3}$Rh$_{4}$Pb$_{13}$. Single unit cell is shown highlighting
the Pb1(Pb2)$_{12}$ icosahedron (light grey).}
\end{figure}

The temperature dependence of the electrical resistivity $\rho$ is
that of a good metal, Fig.~2. $\rho$ decreases from the value of 100
$\mu\Omega\cdot$cm at 300 K to $\sim$ 30 $\mu\Omega\cdot$cm at 2 K.
The inset to Fig.~2 shows a partial superconducting transition at
T$_{c}$=2 K, which we believe is due to trace amounts of
superconducting RhPb$_{2}$~\cite{gendron1962}. Measurements of the
Meissner effect confirmed these conclusions, finding a volume
fraction of less than 1$\%$ for the proposed RhPb$_{2}$ inclusions.

\begin{figure}
\includegraphics[scale=0.5]{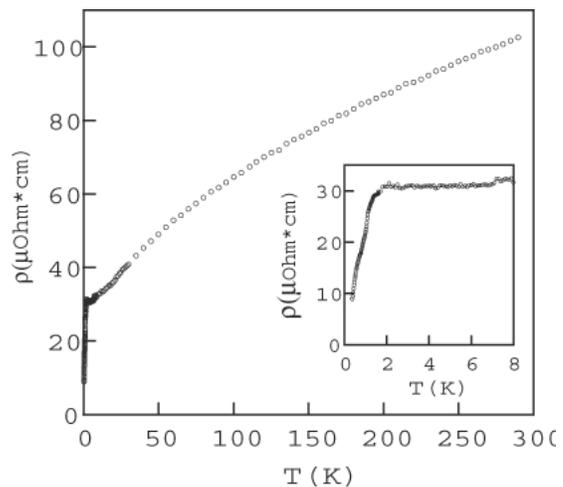}
\caption{\label{fig:epsart} Temperature dependence of the electrical
resistivity of Ce$_{3}$Rh$_{4}$Pb$_{13}$.}
\end{figure}

\begin{figure}
\includegraphics[scale=0.55]{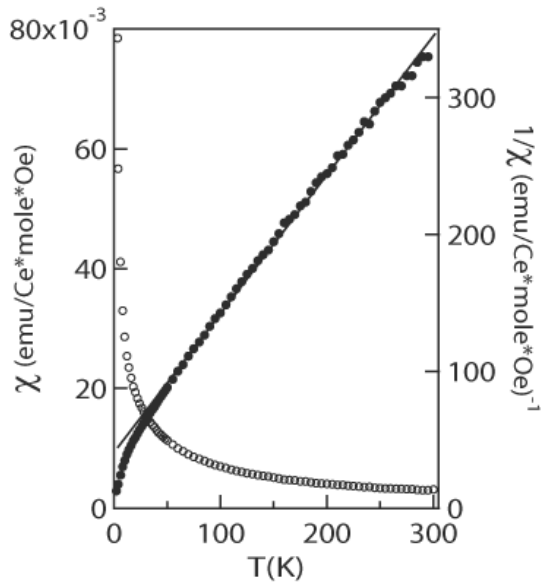}
\caption{\label{fig:epsart} Temperature dependence of the magnetic
susceptibility ($\circ$) and its inverse ($\bullet$) of
Ce$_{3}$Rh$_{4}$Pb$_{13}$ measured in 1000 Oe.}
\end{figure}

Measurements of the magnetic susceptibility found that the Ce ions
in Ce$_{3}$Rh$_{4}$Pb$_{13}$ are well localized. The temperature
dependence of the $\emph{dc}$ magnetic susceptibility $\chi$ and its
inverse are shown in Fig.~3. $\chi$ decreases monotonically with
decreasing temperature, and the inverse of $\chi$ was fitted to a
straight line between 34 K and 300 K, Fig.~3. The fit yielded a
Weiss temperature $\Theta$=-39 K$\pm$0.7 K, suggesting that the
magnetic correlations in Ce$_{3}$Rh$_{4}$Pb$_{13}$ are
antiferromagnetic, and an effective magnetic moment
$\mu_{eff}$=2.6$\pm$0.1 $\mu_{B}$ per Ce ion, close to the 2.54
$\mu_{B}$ expected for a free Ce$^{3+}$ ion. Below $\sim$30 K
1/$\chi$ demonstrates a pronounced downward curvature, as is
characteristic of systems with a substantial crystalline electric
field (CEF) splitting of the ground state of Ce$^{3+}$. $\chi$ is
found to be completely isotropic in the measured temperature range,
which is expected for a cubic system such as
Ce$_{3}$Rh$_{4}$Pb$_{13}$ where Ce ions occupy only equivalent sites
in the unit cell.

Heat capacity measurements confirm that the Ce moments are only
weakly coupled to the conduction electrons. The temperature
dependence of the heat capacity C(T) was measured in zero field and
at temperatures as large as 70 K (Fig.~4a). We have estimated the
phonon contribution to the heat capacity C$_{ph}$ using the Debye
expression and a Debye temperature $\theta_{D}$=163$\pm$10 K.
C$_{ph}$ is subtracted from C(T) in Fig.~4a to isolate the remaining
magnetic and electronic contributions to the heat capacity. The
latter is expected to result in a component of the heat capacity
which is linear in temperature, C$_{el}$=$\gamma$T.

\begin{figure}
\includegraphics[scale=0.45]{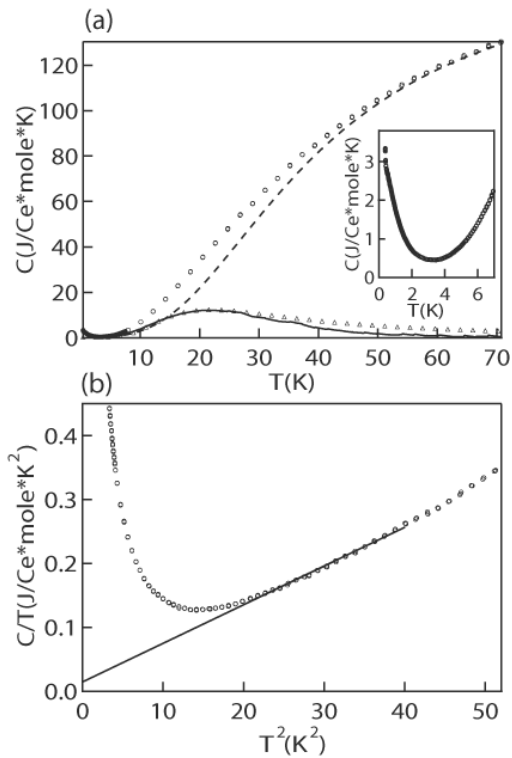}
\caption{\label{fig:epsart} (a)Temperature dependence of the heat
capacity C ($\circ$) of Ce$_{3}$Rh$_{4}$Pb$_{13}$ measured in zero
field. The dashed line is the estimated lattice heat capacity.
$\bigtriangleup$ mark the electronic and magnetic parts of the total
heat capacity. The solid line is the Schottky fit. (b)The electronic
part of the heat capacity C$_{el}$=$\gamma$T is determined from this
plot of C/T=$\gamma$+$\beta$T$^2$.}
\end{figure}

C/T is plotted as a function of T$^{2}$ in Fig.~4b, demonstrating
that the electronic contribution is found only over a narrow range
of temperatures, yielding $\gamma$=15$\pm$2 mJ/moleK$^{2}$. We
conclude that the purely electronic contribution to the heat
capacity is very small, as would be expected for weakly correlated
conduction electrons or alternatively for a low density of
conduction electrons for Ce$_{3}$Rh$_{4}$Pb$_{13}$. In either case,
we conclude that the Kondo temperature for this system is very
small.

The crystal electric field scheme for these localized Ce moments was
established from heat capacity measurements. The CEF of a cubic
symmetry splits the ground state manifold of Ce$^{3+}$ ions with a
total angular momentum $\emph{J}$=5/2, into a doublet and a quartet
~\cite{lea1962}. We suggest that the broad maximum found in
C-C$_{ph}$=C$_{mag}$ near 23 K in Fig.~4a is a Schottky anomaly,
which corresponds to a thermally activated transition from a
$\Gamma_{7}$ doublet ground state to the higher lying $\Gamma_{8}$
quartet, separated by an energy gap of 60 K. Therefore, the full
Ce$^{3+}$ moment can only be found above 60 K. Below $\sim$ 60 K we
expect to find a reduced and temperature dependent effective
magnetic moment.

Given the high crystal symmetry, the relatively close Ce-Ce spacing
and the absence of a measurable Kondo effect, we searched for
evidence that magnetic order occurs in this system. At temperatures
below 3 K the heat capacity demonstrates a monotonic increase,
perhaps indicative of incipient order, shown in the inset of the
Fig.~4a. However, the application of a magnetic field shows that
this is the high temperature side of another broad Schottky peak.
When magnetic field is applied to Ce$_{3}$Rh$_{4}$Pb$_{13}$, the
upturn develops into a broad maximum, which shifts to higher
temperatures with increasing field, Fig.~5. The best fit to the
pronounced maximum found in the 2 T data was obtained assuming the
Schottky-type transition between the levels with the same
degeneracies. A magnetic field of 2 T splits the ground state
doublet into two singlets separated by an energy gap of 2 K.
indicating that the low-lying doublet ground state is itself split,
with an extrapolated energy gap of $\sim$ 1 K in zero field.

\begin{figure}
\includegraphics[scale=0.55]{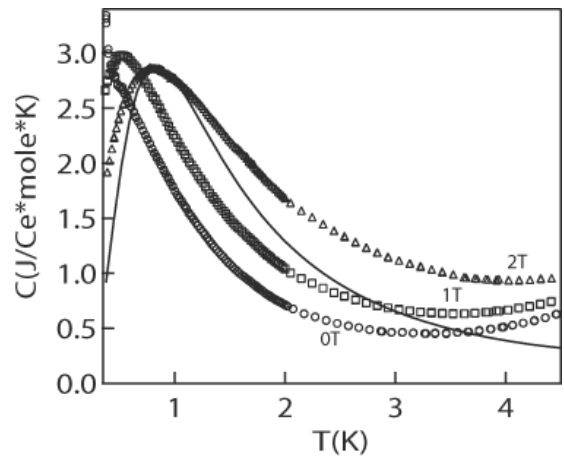}
\caption{\label{fig:epsart} Temperature dependence of the heat
capacity of Ce$_{3}$Rh$_{4}$Pb$_{13}$ measured in 0 T ($\circ$), 1 T
($\square$) and 2 T ($\vartriangle$). Solid line is the Schottky
fit.}
\end{figure}

The temperature dependence of the entropy calculated from the heat
capacity data of Fig.~4a is shown in Fig.~6. The entropy increases
sharply up to T=2 K and almost saturates at the value of $\sim$0.62
Rln2 and increases as the temperature is further increased. The slow
approach of the entropy to the expected value of Rln2 is likely due
in part to the further splitting of the ground doublet, but may also
reveal the onset of a weak Kondo effect, with a Kondo temperature of
no more than 1-2 K.

\begin{figure}
\includegraphics[scale=0.45]{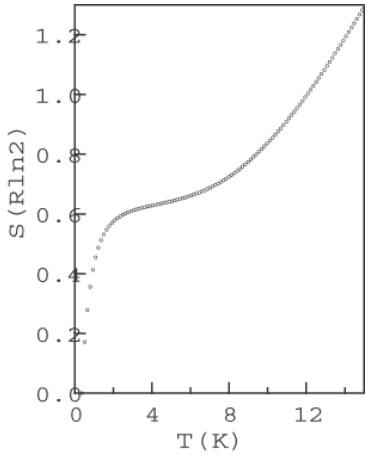}
\caption{\label{fig:epsart} Temperature dependence of the zero field
entropy of Ce$_{3}$Rh$_{4}$Pb$_{13}$.}
\end{figure}

The temperature dependence of the magnetic susceptibility $\chi$ is
in good agreement with the crystal field scheme deduced from the
heat capacity measurements in Ce$_{3}$Rh$_{4}$Pb$_{13}$. Fig.~7
shows the temperature dependence of the effective magnetic moment
defined as $\mu_{eff} (T)=\sqrt{\frac{3k_{B}T\chi}{N}}$. Well above
T=60 K, which is the energy splitting between the ground state
doublet and the excited quartet split by the CEF, $\mu_{eff}$ is
close to the value of 2.54 $\mu_{B}$ expected for a free Ce$^{3+}$
ion. At lower temperatures CEF partially lifts the degeneracy of the
ground state and reduces the total angular momentum $\emph{J}$.
This, in turn, lowers the $\mu_{eff}$ as $\mu_{eff}\propto
\sqrt{\emph{J}(\emph{J}+1})$.
\begin{figure}
\includegraphics[scale=0.55]{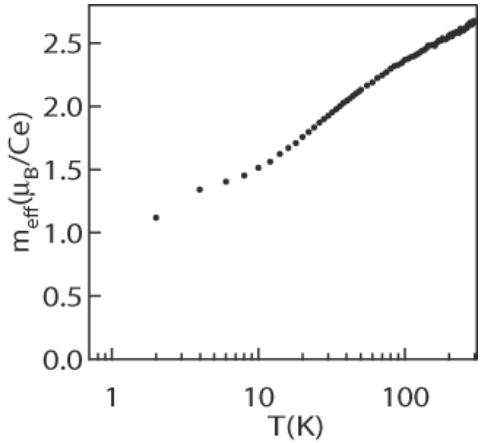}
\caption{\label{fig:epsart} Temperature dependence of the effective
magmetic moment $\mu_{eff} (T)=\sqrt{\frac{3k_{B}T\chi}{N}}$ of
Ce$_{3}$Rh$_{4}$Pb$_{13}$.}
\end{figure}

The derived CEF allows us to calculate the CEF magnetic
susceptibility $\chi_{CEF}$ considering both the Curie and Van-Vleck
terms. We have used the eigenfunctions reported in
Ref.~\cite{lea1962}:

for $\Gamma_{7}$:
$1/\sqrt{6}|\pm5/2\rangle-\sqrt{5/6}|\mp3/2\rangle$

for $\Gamma_{8}$:
$\sqrt{5/6}|\pm5/2\rangle+1/\sqrt{6}|\mp3/2\rangle$ and
$|\pm1/2\rangle$.
\begin{eqnarray}
\chi_{CEF}=
\begin{array}{c}
\frac{(g_{j}\mu_{B})^{2}}{1+2e^{-\Delta/k_{B}T}}
(\frac{\frac{25}{36}+\frac{65}{18}e^{-\Delta/k_{B}T}}{k_{B}T}+\frac{40(1-e^{-\Delta/k_{B}T})}{9\Delta})
\end{array}\;,
\end{eqnarray}
Fig.~8 shows the inverse of $\chi$ and $\chi_{CEF}$. We note that
the agreement between $\chi_{CEF}$ and $\chi$ is reasonable.

\begin{figure}
\includegraphics[scale=0.55]{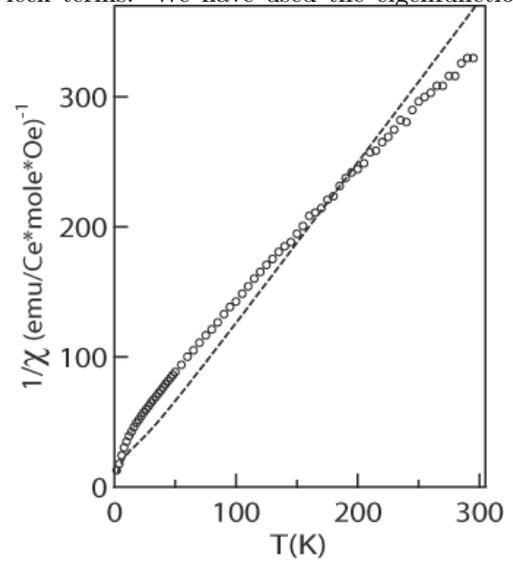}
\caption{\label{fig:epsart} Temperature dependence of the inverse of
the magnetic susceptibility ($\circ$) of Ce$_{3}$Rh$_{4}$Pb$_{13}$
and the inverse of the calculated $\chi_{CEF}$.}
\end{figure}

Taken together, our data reveal Ce$_{3}$Rh$_{4}$Pb$_{13}$ to be an
almost completely localized moment system, with no indication of
magnetic order above 0.35 K, and a similarly small Kondo
temperature. We infer that Ce$_{3}$Rh$_{4}$Pb$_{13}$ most likely
lies at the weakly coupled extreme of the Doniach phase
diagram~\cite{doniach1977}, although we cannot entirely rule out the
possibility that it is very close to the quantum critical point,
where both the magnetic ordering and Kondo temperatures can vanish
simultaneously. Our data suggest that despite high crystal symmetry
and closely spaced Ce moments, Ce$_{3}$Rh$_{4}$Pb$_{13}$ apparently
avoids magnetic order through the successive lifting of the Ce
degeneracy, and the consequent suppression of the Ce moment. It will
be very interesting to extend our measurements to temperatures below
0.4 K to ascertain the ultimate fate of Ce$_{3}$Rh$_{4}$Pb$_{13}$:
magnetic order, Kondo lattice formation, or even superconductivity.

Work at the University of Michigan was performed under grant
NSF-DMR-0405961.

\end{document}